# Tailoring symmetric metallic and magnetic edge states of nanoribbon in semiconductive monolayer PtS$_2$

Ziran Liu[*], Shan Liu, Heyu Zhu and Guanghui Zhou[#]

Department of Physics, Key Laboratory for Low-Dimensional Structures and Quantum Manipulation (Ministry of Education), and Synergetic Innovation Center for Quantum Effects and Applications of Hunan, Hunan Normal University, Changsha 410081, China

*Correspondence to: zrliu@hunnu.edu.cn

#Correspondence to: ghzhou@hunnu.edu.cn

**Abstract:**

Fabrication of atomic scale of metallic wire remains challenging. In present work, a nanoribbon with two parallel symmetric metallic and magnetic edges was designed from semiconductive monolayer PtS$_2$ by employing first-principles calculations based on density functional theory. Edge energy, bonding charge density, band structure and simulated STM of possible edges states of PtS$_2$ were systematically studied. It was found that Pt-terminated edge nanoribbons were the relatively stable metallic and magnetic edge tailored from a noble transition metal dichalcogenides PtS$_2$. The nanoribbon with two atomic metallic wires may have promising application as nano power transmission lines, which at least two lines are needed.

**Graphic abstract**

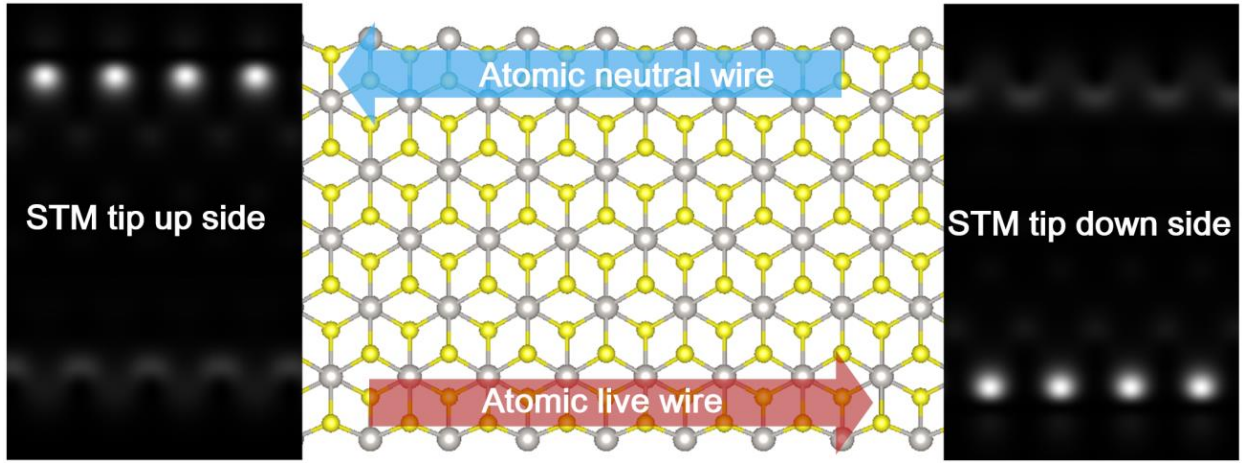

**Symmetric nano power transmission lines, red and blue arrow represent current.**

## 1. Introduction

Traditional one-dimensional metallic wire have been widely studied during the last twenty years using semiconductor heterostructures with length scale of micro meter.[1] Up to date, growth of atomic one-dimensional metallic wire remains challenging, instead scientist try to find metallic edges by tailoring two dimensional semiconductive materials.[2, 3] Such edge states would then be a realization of a one-dimensional metallic wire. Besides graphene,[4] monolayer transition-metal dichalcogenides (TMDs) receive much attention due to their excellent physical properties for both fundamental research and potential applications in electronics, optoelectronics, spintronics and catalysis.[3-11] Recently, in monolayer $MoS_2$ nanoflake, localized metallic states, i.e., one-dimensional metallic wire at the edges have been realized, which may provide a template for studies of quantum phenomena like Aharanov-Bohm effect, persistent currents and weak localization on atomic scale.[11, 12] Controlled bottom-up fabrication of $MoSe_2$ nanoribbons show self-passivation morphological phase transition of the Mo-terminated edges had metallic properties.[13, 14]. However, both atomic structure symmetry of $MoS_2$ and $MoSe_2$ are trigonal (2H phase), which determine the zigzag nanoribbons of those materials should have two different edges (one for Mo-terminated and the other is S-terminated).[3] Mo-terminated edge is metallic and magnetic while S-terminated is not.

Herein we address the question whether a more flexible metallic edge states of zigzag nanoribbons could be realized by tailoring another monolayer material, such as noble TMDs $PtS_2$. Nobel TMDs (10 group) are not our familiar 2D TMDs (6 group).[15] They show extraordinary electronic properties, electrocatalysis and pressure-induced superconductivity,[16-20] attracting much attention recently. What's more interesting is the space group of stable structure of $PtS_2$ is $P\bar{3}m1$, with the Pt in octahedral coordination (2H-$MoS_2$ is $P\bar{6}m2$, where Mo site is in a trigonal prism coordination).[21] Such structure of 2D $PtS_2$ is convenient for tailoring suitable edge states of nanoribbons or nanoflakes with desired one-dimensional metallic edges. In this paper, we employ first principles calculations within density functional theory (DFT) to explore systematically all possible edges states of zigzag and armchair $PtS_2$ nanoribbons.

## 2. Methodology

First principles calculations were based on density functional theory (DFT) implemented in the Vienna Ab initio Simulation Package (VASP) [22-24] with projector-augmented wave (PAW) potential [25]. The generalized gradient approximation (GGA) with the exchange-correlation functional of Perdew-Burke-Ernzerhof (PBE) [26] was employed. Convergence tests indicated that 360 eV was a sufficient cutoff (Higher than 1.3*default value) for PAW potential to achieve high precision in all the zigzag and armchair nanoribbons supercells calculations. The global break condition for the electronic self-consistency is chosen as $10^{-5}$ eV per supercell. Due to the ferromagnetic nature of Pt and Mo, all related calculations were performed using spin polarized approximation unless otherwise noted. All atoms on edges and on center partial of the nanoribbons are fully optimized during calculations until the residual forces were less than 0.01 eV/Å by using conjugate gradient (CG) algorithm. STM images were simulated in constant-current mode at different bias voltages by using Tersoff-Hamann approximation, which illustrate that the tunneling current is proportional to local density of states (LDOS) of the sample.[27] The crystal structure and charge density are plotted by VESTA.[28]

## 3. Results and discussion

For electron properties of monolayer $PtS_2$. It is a semiconductor with indirect bandgap of 1.60 eV from experiment.[29] Our DFT calculation indicates an indirect bandgap of 1.81 eV as shown in figure 1, which well consistent with previous simulation work of 1.80 eV.[19] For the structure of nanoribbons of $PtS_2$, there are three possible zigzag nanoribbon edges could be cut from $PtS_2$, while the armchair nanoribbon have only one choice, as shown in figure 2 (a). To avoid the interaction between the two edges, the minimum width of nanoribbon used in the calculations was sufficiently large as 18.11Å (Pt-S edge supercell).

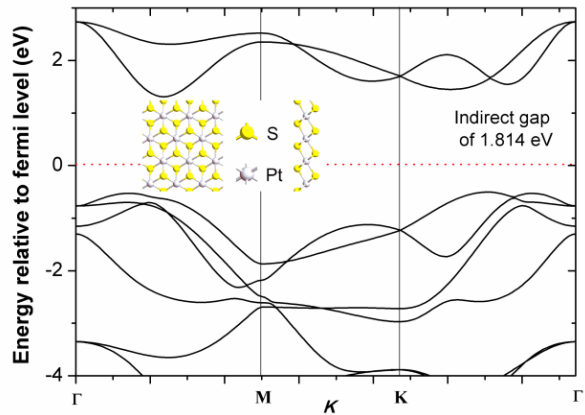

Figure 1. The band structure of monolayer PtS2.

The stability of PtS₂ nanoribbons are quite important since it determine whether the nanostructure with such edges could be realized experimentally. The stabilities of nanoribbons were determined by the stabilities of the edges. We built Pt-S, S-Pt, S-S and armchair edge totally four nanoribbon supercells, as shown in figure 2 (b-e). The edge at one side is equal to the opposite side, though they might not sharp symmetrical. In order to determine the stabilities of the edges, we first relax all the atoms in the four supercells. During the relaxation, Pt and S atoms along all the four edges supercells are slightly draw back, while the position of atoms at center remain unchanged. Take Pt-S edge nanoribbon supercell for example, two of three Pt-S bonding along edge relaxed to be shorter, from 2.4 Å to 2.2 Å, resulting in two angle of S-Pt-S to be larger, from 96° to 107° and 84° to 93° separately. The relaxation reveal that the zigzag and armchair nanoribbons of PtS₂ are mechanically stable, because atoms along edges didn't obviously move during relaxing.

The edge energy was employed to determine the stabilities of the nanoribbons as well. It was successfully used to elucidate the stabilities of MoS₂ edges.[30] The edge energy σ of PtS₂ here could be defined as:

$$\sigma = \frac{1}{N * 2L}\left(E - N_{Pt} * \mu_{PtS_2} - \Delta N * \mu_S\right)$$

where $\Delta N = N_S - 2N_{Pt}$ and $L$ represent the excess of the S atoms at the edges and the length of the edge respectively. $E$ and $N$ are the total energy and total number of atoms of the supercell. For example, $\Delta N = 0$ implying that the edge energy is independent of chemical potential, while $\Delta N \neq 0$, the edge energies are linear function of chemical potential of sulfur atom. Note this could be equal in terms of Pt chemical potential since $\mu_{PtS_2} = \mu_{Pt} + 2 \cdot \mu_S$. For the edge energy related to stability, those with lowest curve are most stable.

The edge energies of all the possible nanoribbons were summarized in figure 4. Present results demonstrate that the most stable PtS$_2$ edge are Pt-S for $\mu_S < -5.8\ eV$, armchair edge for $-5.8\ \text{eV} < \mu_S < -3.2\ eV$ and S-S edge for $\mu_S > -3.2\ eV$. This can be understood by simply relating high S chemical potential to S-rich stable edges, and low one to S-poor stable edges (or Pt-rich edge).

Chemical bonding between Pt and S are essential for understanding the stability of the monolayer PtS$_2$ and its edges [31, 32]. The properties of chemical bonding could be studied by calculating the bonding charge densities, which defined as the charge transfer upon the bonding formation between atoms, i.e., the valence charge density from self-consistent DFT calculations minus the superposition charge density of all the involved atoms. In figure 5 (a) and (b), obvious blue isosurfaces are shown along the edge, which indicate there are dangling bonds along the armchair and S-Pt edge nanoribbons. For S-S edge nanoribbon in figure 5 (c), the S atoms at the most outside being isolated, this always not good for a stable S-S edge nanoribbon.[30] Pt-S edge seems don't have too many dangling bonds or isolated atoms, should be a promising stable edge states.

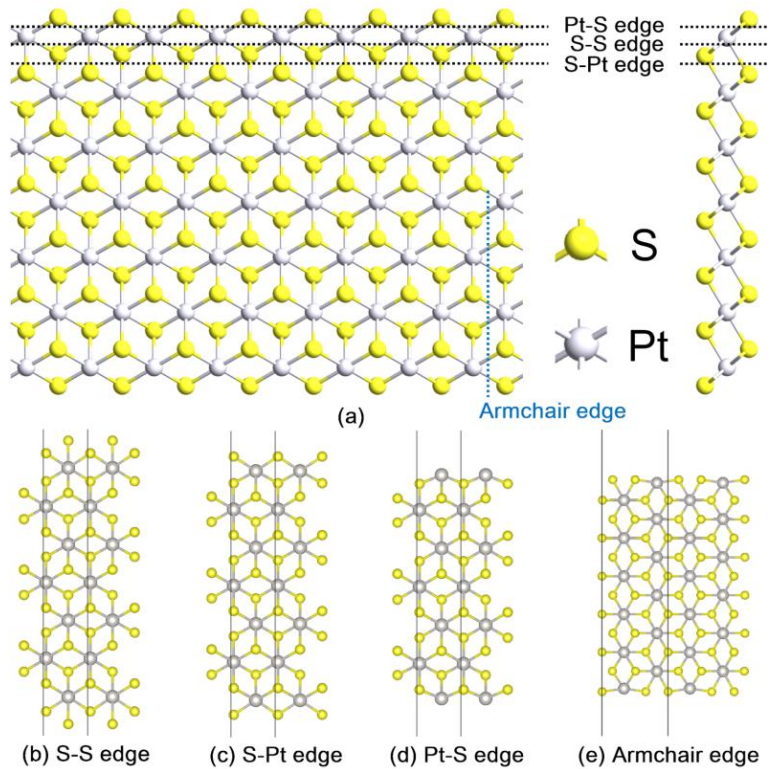

Figure 2. Schematic drawing of three zigzag edges and one armchair edge of PtS$_2$ nanoribbon, top- and side-views. (a) Pt-S edge, S-S edge and S-Pt edge are the three possible termination of zigzag nanoribbon being tailored along the dot line; (b-e) Relaxed S-S edge, S-Pt, Pt-S and armchair edge supercell separately.

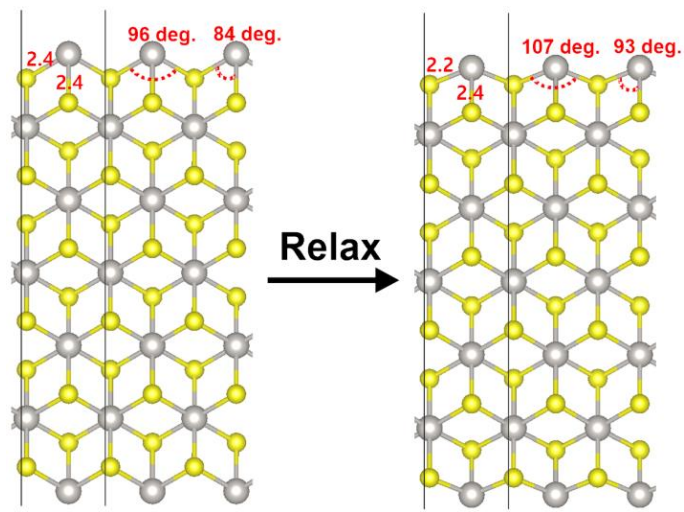

Figure 3. Pt-S edge supercell. Left is the initial structure, right is the relaxed one.

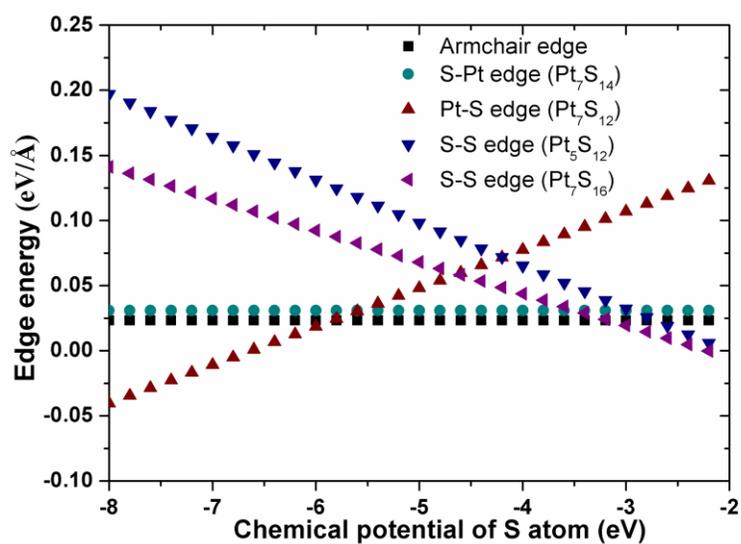

Figure 4. The edge energies of the four types of edges as function of sulfur chemical potential in zigzag and armchair nanoribbon of $PtS_2$.

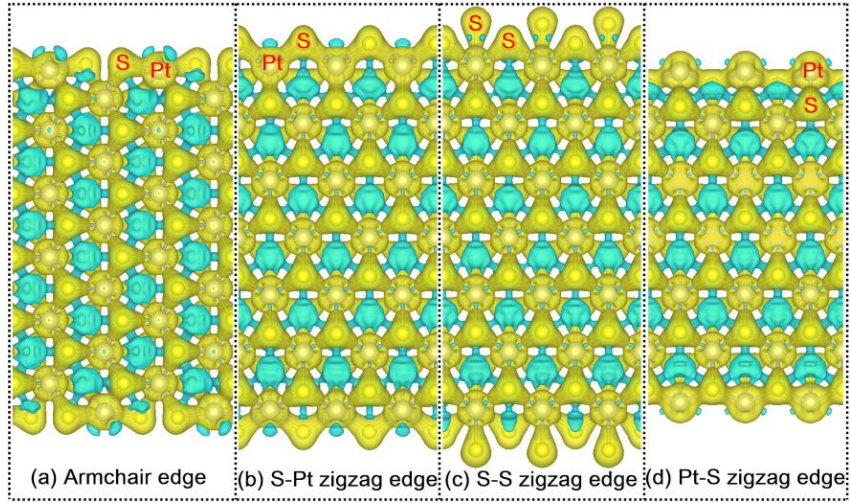

Figure 5. Isosurfaces of differential charge density contours of armchair (a) and zigzag (b-d) edges. The same isosurfaces levels of 0.008 eV$^{-3}$ were used to draw the plots. Blue zones represents accumulation of electron, while yellow region represents loss of electron.

Figure out the effects from edges on electronic states in the nanoribbons are crucial to guide the tailoring of 2D PtS$_2$. The band structure is an important and useful way. From figure 6, it is demonstrated that the semiconductive monolayer PtS$_2$ still act as semiconductor though being tailoring into nanoribbons with armchair, S-Pt and S-S edges.

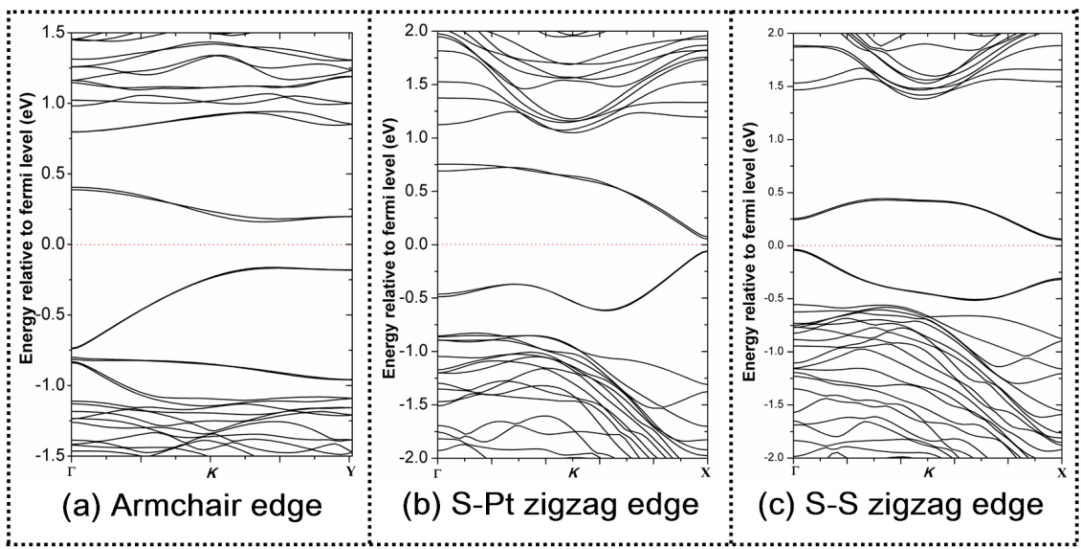

Figure 6. The band structure of nanoribbons with (a) armchair edge, (b) S-Pt edge and (c) S-S edge.

Interestingly, semiconductive nanoribbon of $PtS_2$ with Pt-S edge are different from the above three ones, it has metallic and magnetic properties, as shown in figure 7. From the band structure, it is clear that two spin up and two spin down bands are crossing the fermi level, which indicate a metallic properties of the nanoribbon. We also find that the center eight bands (including 4 spin up and 4 spin down bands) near or cross the fermi level are drawn by using edge atoms without exception. In order to show a real space image of electron states with energy around the fermi level, a partial charge density calculations were carried out. From figure 7 (right), we may easily understand that those electron states are coming from edges atoms in real space, rather than from center atoms.

Which type of atom and what orbital mainly contribute to the metallic state of nanoribbon edges? To address this question, a density of states calculation was needed. Nearby fermi level, the density of states from all atoms and that from edges atoms are almost equal, as shown in figure 8. This is another proof that the metallic states come from two parallel edges. More details calculations show that the metallic states mainly contributed from Pt atoms located at edges. We also check all the possible contribution from all kind of orbits. It is found that the main orbit of Pt is 3d. Certainly, there are contribution from S atom of 2p orbit, but they are relatively small. Our calculations also show the Pt 3d orbit have hybrid with S 2p orbit, which demonstrate that the Pt and S have formed strong bonds.

For the magnetism of the nanoribbons. We do a spin polarized and spin unpolarized calculations for the above four supercells. Then we check the total energy and magnetism. It was found that only for Pt-S edge nanoribbons, the total energy from spin polarized and spin unpolarized calculations are different, the other three nanoribbons are have the same energies, and thus no magnetism. The energy difference of Pt-S edge nanoribbons -72 meV per atom. The magnetic moment per unit cell **M** is 1.609 $\mu$B.

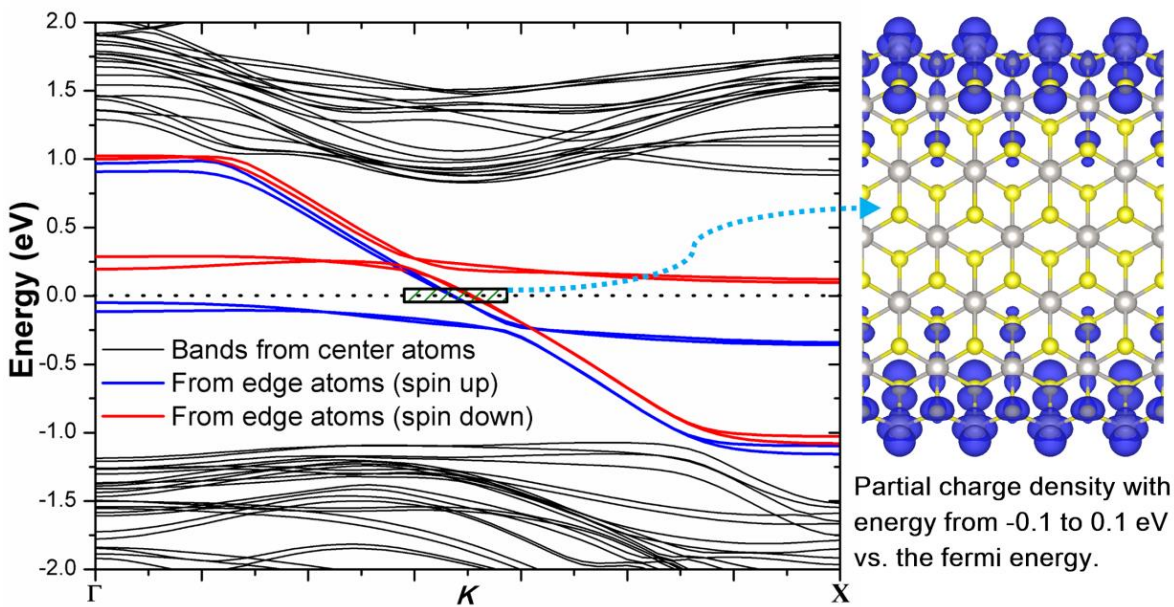

Figure 7. The band structure and partial charge density of nanoribbons with Pt-S edges. There are 4 red and 4 blue bands around the fermi level. Here edge atoms including two parallel edges, one of them have 2 Pt and 2 S atoms together.

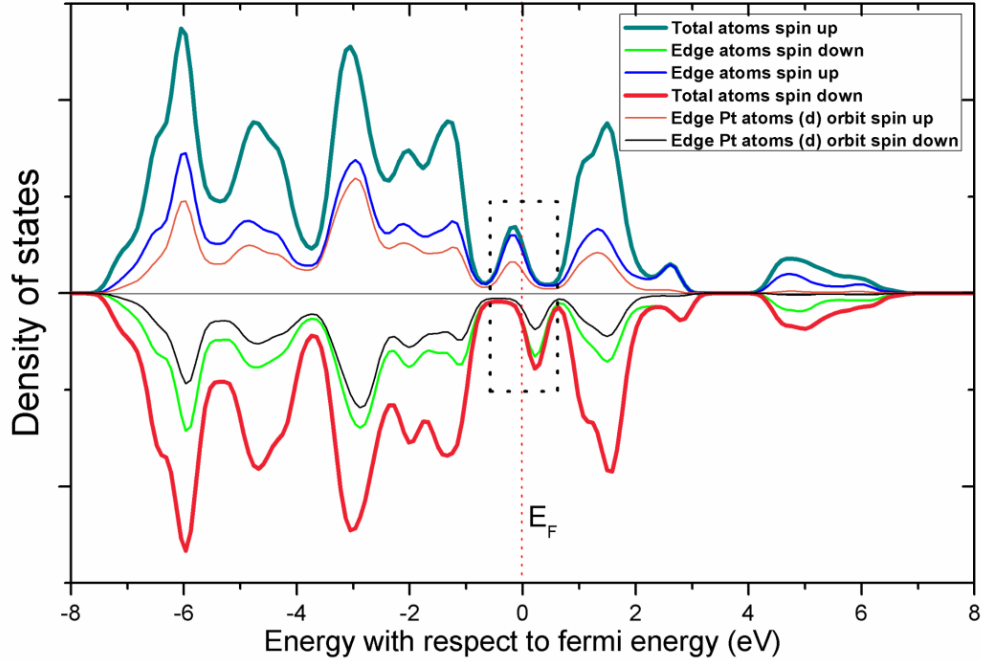

Figure 8. The density of states of nanoribbons with Pt-S edges.

The distinctly different patterns of local density of states (LDOS) could provide an important guide for differentiating metallic edges once formed in experiment.[27] So we have simulated scanning tunneling microscopy (STM) image of Pt-S edge nanoribbon and have taken that of $MoS_2$ zigzag edge nanoribbon as a reference. Like $MoS_2$, Pt-S edge nanoribbon show promising properties of one-dimensional metallic wire. What different from $MoS_2$ nanoribbon is that there are white spots composing only one line of conductive nanowire, while in $MoS_2$, there are two lines of that, as shown in figure 9 (a, b and d, e). At last, we do a STM simulation on the opposite surface, as shown in figure 9 (c, f). It is interesting that both Pt-S edge nanoribbons are metallic, but $MoS_2$ nanoribbon only have one edge terminated with Mo. The equal parallel edges of metallic wire have promising application in realization of quantum effect. Such as atomic power transmission lines, only one Pt-S edges of nanoribbon is enough. According to symmetry of $PtS_2$, it is also possible to fabricate a $PtS_2$ triangle nanoflake, just as in $MoS_2$.[11]

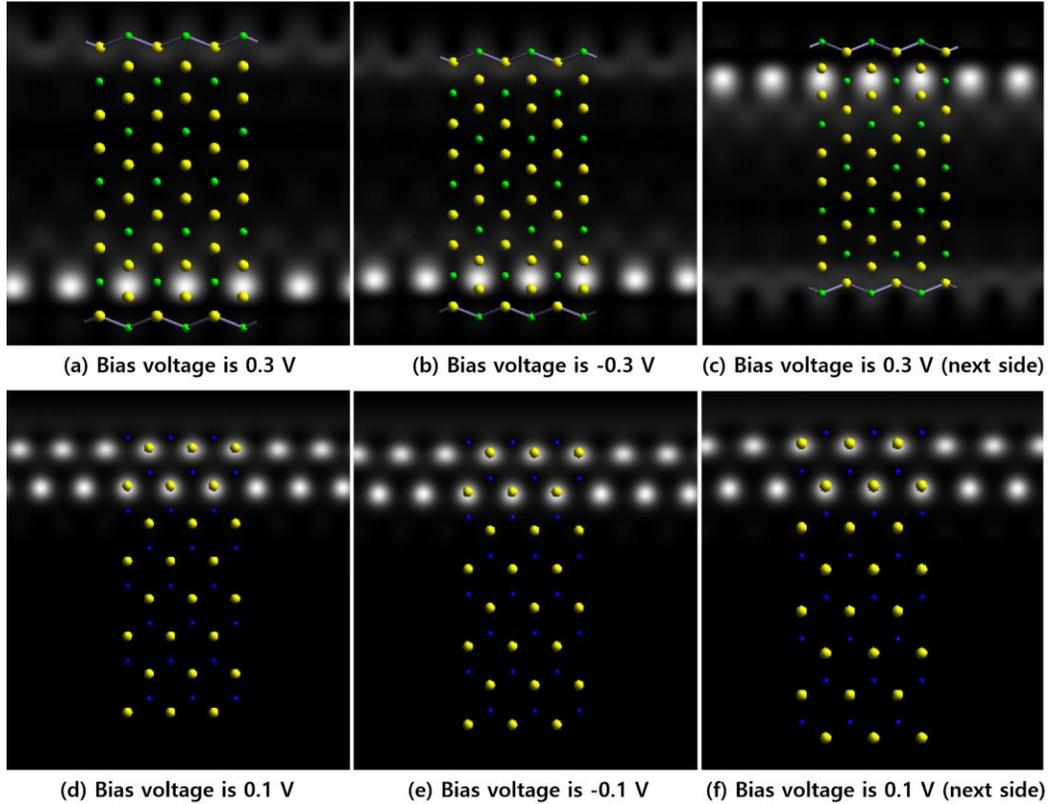

Figure 9. Simulated constant-height STM image of zigzag nanoribbons of $PtS_2$ (a-c) and $MoS_2$ (d-f) at different bias voltages. A ball and stick model of nanoribbons is superposed onto the simulated STM images. The tip plane is above the surface atoms about 1 Å. (c) and (f) are the STM images of next side of zigzag nanoribbons of $PtS_2$ and $MoS_2$ separately. Yellow ball is S atom, green ball is Pt atom and blue ball is Mo atom.

## 4. Summary

We have systematically investigated the edge states of $PtS_2$ nanoribbons by using first-principles calculations. The atoms along Pt-S edge displace slightly during the relaxation. The energetic favorable Pt-S terminated zigzag edges have relatively weak dangling bonds and less isolated atoms along the edge. Those stabilities indicate that Pt-S zigzag edge of nanoribbon may be fabricated in experiment. Metallic and magnetic properties of the Pt-S zigzag nanoribbon mainly determined by edges atoms. To be more detailed, the metallic and magnetic properties come from *d* orbit of Pt atom at the edges. Simulated STM show Pt-S zigzag nanoribbon could have two parallel symmetric metallic wire in one nanoribbon, which have promising application for atomic power transmission lines.

**Acknowledgments**

The authors are grateful for financial support from National Natural Science Foundation of China (General program 51671086), Hunan Provincial Natural Science Foundation of China (2016JJ3089) and excellent youth fund of Hunan Provincial Education Department (16B151). Appreciate helpful discussion with W. K. Wang.